\documentstyle[epsfig]{hep99}
\begin{document}

\title{Systematic Study of Hadronic Observables in Nucleus Nucleus Collisions at CERN SPS}

\author{R. Ganz$^1$ for the NA49 collaboration}

\address{$^1$ Max-Planck-Institut f\"ur Physik M\"unchen, F\"ohringer Ring 6, 
D-80805 M\"unchen, Germany\\[3pt]
e-mail: {\tt Rudolf.Ganz@cern.ch}}

\abstract{The possible transition of nuclear matter to a deconfined phase 
in relativistic nucleus-nucleus collisions is explored by a systematic variation of 
the collision system by means of system size, beam energy and centrality of the collision
all measured within the same experiment NA49 at the CERN SPS. 
The investigation takes advantage of the large number of hadronic observables
covered over a wide phase space by the acceptance of the NA49 TPCs.} 

\maketitle

\section{Introduction}
The aim of studying nucleus-nucleus collisions at relativistic beam energies
is the potential creation of a new state of matter, the Quark Gluon Plasma, 
in which quarks are no longer confined to a single nucleon but can move quasi-freely over a
larger volume. The highly dense environment, an essential for the occurance of this state,  
might be established for a short period of time in collisions of two heavy nuclei
at relativistic beam energies at the CERN SPS.
Because so far no signature has been put forward, 
which would uniquely identify the transition 
to such a state simply by its mere observation,
only a systematic study can reveal the onset of such a new phenomenon.
The NA49 collaboration at CERN SPS has undertaken such a systematic approach by  
a variation of the initial state of the collision in terms of collision systems
(p+p, p+Pb and Pb+Pb), different centralities and two beam energies (40 AGeV and 158 AGeV).
Due to the versatility of NA49\cite{NIM99} in all these systems 
a large number of observables has been surveyed 
over wide phase space (see \cite{Sikler} and ref. therein), 
all within the same experimental setup.
Here we present only a few aspects focussing on identified charged
particle spectra and two particle correlations.

\section{Identified particle spectra}
In NA49 two methods of identifying particles are utilized. 
The more stringent method utilizes
a time-of-flight measurement by four scintillator walls 
over a flight-path of 13~m. 
Since in this case the coverage of phase space is rather limited 
the results presented here are based on 
the second method, which exploits the specific energy loss 
$dE/dx$ (resolution typically 4\%) 
deposited by charged particles in the gas of the four large volume time 
projections chambers of NA49\cite{NIM99}.

\begin{figure}
\begin{center}
\epsfig{file=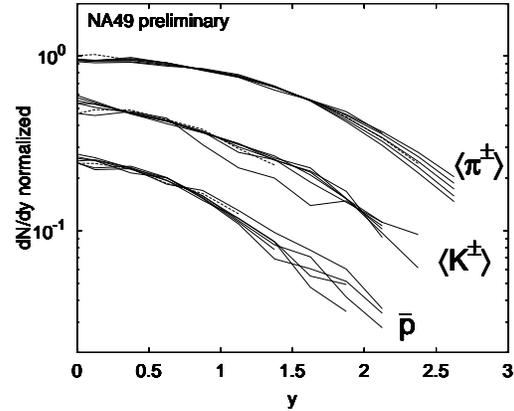,width=0.9\linewidth}
\caption{Shape of the rapidity distribution of average charged pions, kaons
and anti-protons. The results are from p+p collisions (dashed line) and Pb+Pb
collision (solid lines) in six bins of centrality and are normalized at $y=0$.}
\label{fig:long}
\end{center}
\end{figure}
Figure~\ref{fig:long} shows a comparison of the shapes of the rapidity distributions
for pions ($<\pi^++\pi^-$$>/2$), 
kaons ($<$K$^++$K$^->$$/2$), 
and anti-protons 
measured in p+p collisions as well as in Pb+Pb collisions at six different
bins of centrality (average impact parameters $<b> \approx$ 2.6, 4.6, 5.7, 7.0, 8.5 and 10.5~fm) 
all at the incident beam energy of 158~AGeV.
It appears that the shapes are almost identical over 
this large variety of collision systems even 
though the change in system volume and geometry is rather sizeable.

\begin{figure}
\begin{center}
\epsfig{file=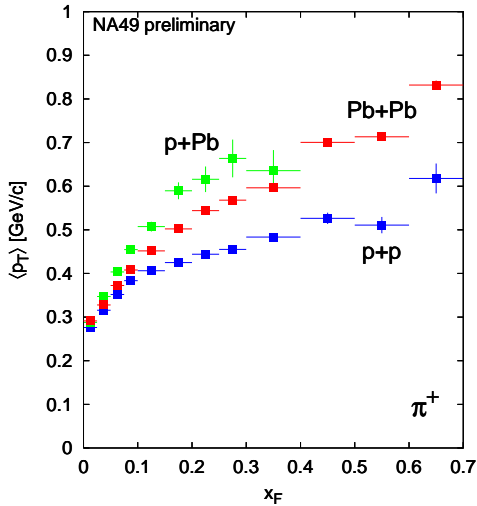,width=0.8\linewidth}
\caption{Average transverse momentum in dependence on Feynman-x $x_F$
for $\pi^+$ in p+p, central p+Pb and central Pb+Pb collsions.}\label{fig:seagull}
\end{center}
\end{figure}
\begin{figure}
\begin{center}
\epsfig{file=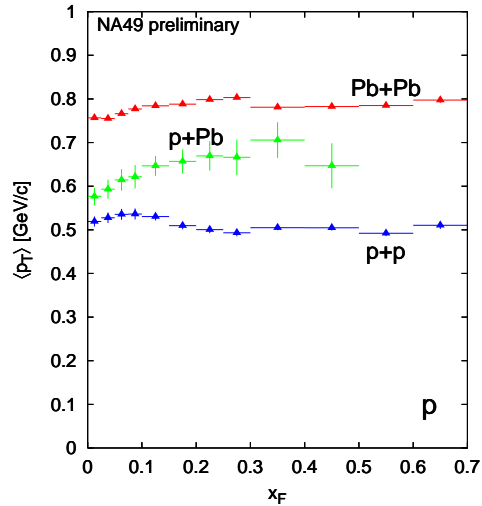,width=0.8\linewidth}
\caption{Same representation as previous figure
but shown for protons instead.}\label{fig:noseagull} 
\end{center}
\end{figure}

Turning attention to the transverse momentum distribution, 
the first question one might ask is, 
whether it is independent from the longitudinal component. 
This is addressed in figures~\ref{fig:seagull} and \ref{fig:noseagull} 
by plotting the average transverse momentum $<$$p_T$$>$ versus 
the longitudinal momentum component, the latter being represented 
by the x-Feynman $x_F$ variable. 
For pions the seagull-shaped wing of figure~\ref{fig:seagull} 
contradicts the independence of the two\footnote{A fact not to be necleted 
when extrapolating results from narrow acceptance experiments into full phase space.}.
This might be due to the contribution of low-$p_T$ pions from resonance decays
at mid-rapidity, whereas a direct production rules at more forward rapidities.

For proton spectra the situation is changed. 
Here, with the exception of the asymmetric projectile-target system p+Pb,
the assumption of an independence between longitudinal and transeverse momentum
is well supported by an almost constant $<$p$_T$$>$ value 
for regions of rapidities (figure~\ref{fig:noseagull}). 
To interpret this one should keep in mind that in this energy regime 
the proton yield is still dominated by
the number of incoming projectile and target nucleons.
The overall increase in $<$p$_T$$>$ from p+p to Pb+Pb
might be attributed to the increased transfer of longitudinal 
momentum into transversal degrees of freedom due to
multiple collisions.

\section{Two particle correlations}

\begin{figure}
\begin{center}
\epsfig{file=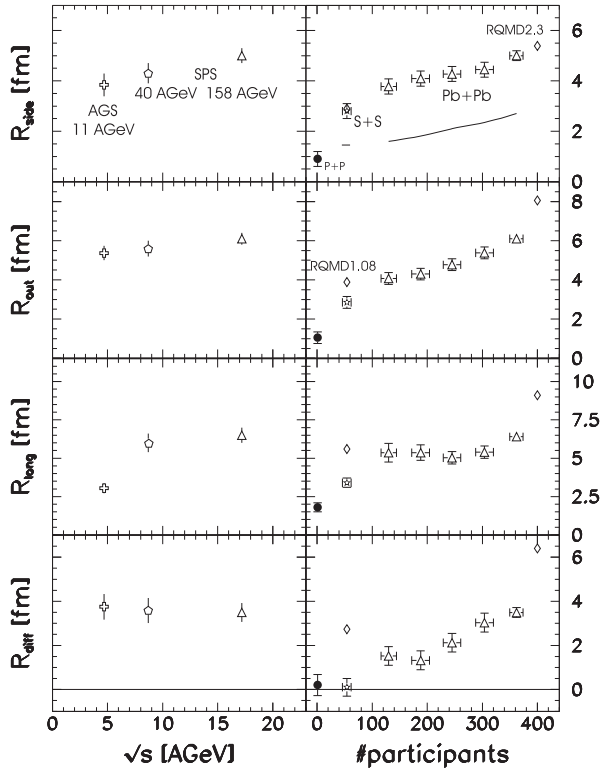,width=\linewidth}
\caption{{\it Left column:} Dependence of radii on the center of mass energy. 
The 4.7 AGeV points are AGS Au$+$Au data from E866 $[6]$.
The 8.7 AGeV radii are our very preliminary 40~AGeV Pb$+$Pb results 
and the 17.2 AGeV points are for central 158 AGeV Pb$+$Pb.
{\it Right column:} Dependence of the radii on number of participants 
in the collision.
The $\bigtriangleup$ correspond to different centralities in 
158 AGeV Pb$+$Pb. 
Symbol {\large$\star$} is for central 200 AGeV S$+$S $[5]$
and {\large$\bullet$} is for the 158 AGeV p$+$p result. The $\diamond$ is  
the RQMD 1.08 result for central S$+$S 
and the RQMD 2.3 result for central Pb$+$Pb. 
The line in $R_{\mathrm{side}}$ 
corresponds to the geometrical transverse size of the overlap region
of the two colliding nuclei.
($<y_{\mathrm{\pi\pi}}>\approx 4.2$ and 
$<k_{\mathrm{t}}>\approx0.12$~GeV$/$c .)}\label{fig:hbt} 
\end{center}
\end{figure}

Two particle correlations, here by means of the quantum mechanical interference 
of two identical pions, 
offer the opportunity to study the space-time evolution
of heavy ion collisions in a direct way.
With respect to the observation of a deconfined phase, this might be important
since a transition should be accompanied by an increase in entropy, 
which in the final state of the collision
should emerge as an increase in overall multiplicty or/and 
as an increase in the systems space-time extent
compared to the case when no such transition has occured.

In the right column of figure~\ref{fig:hbt} the dependence 
of the fitted radius parameters on
the number $n_{\mathrm{part}}$ of nucleons participating in the collision
is shown for Pb$+$Pb collisions at different centralities
compared to NA49 results from p$+$p reactions and to the NA35 results \cite{NA35SS}  
on S$+$S at 200~AGeV.
The transverse radii $R_{\mathrm{side}}$ and $R_{\mathrm{out}}$ 
grow continously whereas $R_{\mathrm{long}}$ appears constant
when increasing $n_{\mathrm{part}}$.
$R^2_{\mathrm{diff}}= R^2_{\mathrm{out}} - R^2_{\mathrm{side}}$, 
being sensitive to the duration of the particle emission phase,
is positive over the full range and $R_{\mathrm{diff}}$ 
grows linearly with $n_{\mathrm{part}}$. 
When comparing the measured radii with those extracted from 
the RQMD model, similar to the findings of NA35, 
a good agreement is found for $R_{\mathrm{side}}$, but both versions
of the model overestimate $R_{\mathrm{out}}$ and $R_{\mathrm{long}}$. 
It is instructive to compare $R_{\mathrm{side}}$ with an estimate 
of the geometrical transverse size of the overlap region 
in the collsions (line\footnote{The line in $R_{\mathrm{side}}$ 
corresponds to the average distance of points
in the overlap region of two spheres of radius $1.16 A^{1/3}$ 
offset by the impact paramter from the center of the the collision,
projected into a randomly oriented reaction plane.}
in figure~\ref{fig:hbt}) which demonstrates 
that the interferometric radii reflect the late freeze-out stage
which is preceeded by a strong expansion of the dense collision
zone produced shortly after the collision \cite{NA49exp}.
The left column of figure~\ref{fig:hbt} shows the collision energy
dependence of the radius parameters displaying results for
the highest AGS energy at 11.6 AGeV, the SPS measurements at 158 AGeV and the 
(very preliminary) results from the 40~AGeV commissioning run 
of NA49 in 1998. The 40~AGeV and the 158~AGeV results correspond to 
$\frac{<y_{\mathrm{\pi\pi}}> - y_{\mathrm{cms}}}{y_{\mathrm{cms}}}
\approx 1.0$ whereas the AGS measurement \cite{E866} 
covers mid-rapidity. 
The comparison is justified by the observed slow
variation of the the radii with $y_{\mathrm{\pi\pi}}$ in NA49.
From AGS to the highest SPS energy $R_{\mathrm{long}}$ 
more than doubles 
whereas the transverse radii 
$R_{\mathrm{out}}$ and $R_{\mathrm{side}}$ and the derived
temporal component $R_{\mathrm{diff}}$ 
stay almost constant.  

\section{Summary}
A systematic study of hadronic observables in heavy ion collisions
has been carried out by the NA49 experiment.
In none of the discussed variables a sudden change has been observed, 
which would indicate a sharp QGP phase transition.
In particular the longitudinal shapes of the produced particle
spectra seem to be unchanged coming from p+p moving
towards the most central Pb+Pb collisions.
The interferometric radii follow smooth trends 
when increasing centrality, when changing the collision system or 
(with the exception of $R_{\mathrm{long}}$)
when increasing the incident beam energy from the AGS to SPS.

\end{document}